\documentclass[a4paper]{article}
\usepackage{enumitem, amsmath}
\usepackage{amsmath,amssymb,graphicx,dblfloatfix,mathtools}
\usepackage{amssymb,relsize}
\usepackage{color, xcolor}
\usepackage[hyperfootnotes=false]{hyperref}
\usepackage[T1]{fontenc}
\newcommand*\email{\familydefault{\ttdefault}}
\usepackage[skip=0pt]{caption}
\newcommand{\bigF}{\scalebox{1.2}{$\mathcal{F}$}}

\usepackage{INTERSPEECH2022}

\title{Joint Neural AEC and Beamforming with Double-Talk Detection}
\name{Vinay Kothapally$^\ast$\thanks{This work was done while V. Kothapally was an intern at Tencent.}, Yong Xu$^\dagger$, Meng Yu$^\dagger$, Shi-Xiong Zhang$^\dagger$, Dong Yu$^\dagger$}
\address{$^\ast$Center for Robust Speech Systems (CRSS), The University of Texas at Dallas, TX, USA\\$^\dagger$Tencent AI Lab, Bellevue, WA, USA}
\email{$^\ast$vinay.kothapally@utdallas.edu, $^\dagger$\{lucayongxu, raymondmyu, auszhang,  dyu\}@tencent.com}

\begin{document}

\maketitle
\begin{abstract}
Acoustic echo cancellation (AEC) in full-duplex communication systems eliminates acoustic feedback. However, nonlinear distortions induced by audio devices, background noise, reverberation, and double-talk reduce the efficiency of conventional AEC systems. Several hybrid AEC models were proposed to address this, which use deep learning models to suppress residual echo from standard adaptive filtering. This paper proposes deep learning-based joint AEC and beamforming model (JAECBF) building on our previous self-attentive recurrent neural network (RNN) beamformer. The proposed network consists of two modules: (i) multi-channel neural-AEC, and (ii) joint AEC-RNN beamformer with a double-talk detection (DTD) that computes time-frequency (T-F) beamforming weights. We train the proposed model in an end-to-end approach to eliminate background noise and echoes from far-end audio devices, which include nonlinear distortions. From experimental evaluations, we find the proposed network outperforms other multi-channel AEC and denoising systems in terms of speech recognition rate and overall speech quality.


\end{abstract}
\noindent\textbf{Index Terms}: Neural AEC, Neural beamforming, Speech enhancement, Deep neural networks

\section{Introduction}
\label{sec:intro}
With an increasing demand for hands-free communication between speakers in two distant locations (far-end and near-end), effective communication necessitates high-quality audio transmission \cite{wangHandsfree,AEC_01}. On the other hand, far-end speakers receive modified versions of their speech as feedback (far-end echo) due to acoustic coupling between the loudspeaker and the microphone locations at the near-end speaker, resulting in reduced speech intelligibility \cite{benestyAEC, hanslerAEC}. To improve the overall communication quality, an AEC system aims to remove far-end speech captured by the microphone at the near-end while preserving speech from the near-end speaker before transmission. Many AEC systems based on digital signal processing (DSP) have used linear and nonlinear adaptive filters to address this issue for more than two decades \cite{GrantNLMS,FastNLMS,FDAEC,BlockAdapFilters,MultiDelayFDAP,nonlinearRES01,wRLS,dtd_mcAEC}. Nonetheless, in practical scenarios involving nonlinear distortions from loudspeakers, the presence of reverberation and background noise at the near-end, and double-talk conditions, their performance in suppressing only far-end speech was insufficient.

\noindent Recent advances in deep learning have shown the potential to improve the performance of many speech processing systems. As a result, hybrid systems combining traditional adaptive filters and DNNs \cite{NLAEC_01,NLAEC_02,NLAEC_04,NLAEC_05,NLAEC_07} have been proposed to address nonlinear distortions from loudspeakers by suppressing residual far-end speech from the adaptive filter output. For example, Speex and WebRTC \cite{Speex1,Speex2} have been combined with RNNs in \cite{NLAEC_03}. Furthermore, multi-task networks were used to design AEC systems with the secondary task of detecting double-talk scenarios in order to avoid suppressing near-end speech in double-talk scenarios \cite{DTD_01,DTD_02,DTD_03}. Later, advanced networks such as complex-valued DNNs \cite{complex_01,AEC_DCNN_01,DNN_02}, Long Short Term Memory networks (LSTM), and multi-head self-attention \cite{DTD_01,attn_02} were used to develop AEC systems to also  compensate for the time lag between far-end speech and microphone captured signal alongside handling nonlinear distortions and double-talk. Early attempts at AEC for multi-channel speech systems included using single-channel AEC on individual microphones, followed by traditional beamforming techniques \cite{combine_adap_beamforming_01, combine_adap_beamforming_02}. Later, end-to-end DNN-based approaches were proposed for multi-microphone AEC systems  \cite{Multi_Channel,dnn_JAB1}.

This paper proposes a novel deep learning-based framework for joint AEC and beamforming. Our main contributions towards the proposed model are summarized as follows: (i) In contrast to DNN-based "black-box" approaches, we include explicit modules for AEC and beamforming, (ii) cross-channel correlation between the echo and multi-channel signals are considered in our neural AEC and beamforming modules to find a better solution for joint AEC and beamforming, (iii) we extend our recently proposed generalized RNN beamformer (GRNNBF) to a spatio-temporal joint AEC-RNN beamformer (JAECBF) capable of learning a better beamforming solution using microphone array signals and processed signals from multi-channel neural-AEC, and (iv) we propose a double-talk detection (DTD) module based on the multi-head attention \cite{yong1_ATT}, to leverage DTD predictions to suppress far-end residuals. 




\section{Problem Definition}
\label{sec:signalmodel}
We consider the problem of enhancing near-end speech captured by $M$-microphones in presence of reverberation, far-end echoes from the loudspeaker, and background noise. Let $s(t)$ and $x(t)$ represent the clean speech from near-end and far-end speakers respectively. The signals captured by an $M$-channel microphone array, $\mathbf{d}(t)$ (termed as ``mixture") at time `$t$' can be represented as, 

\begin{equation} 
\begin{aligned}[b]
\mathbf{d}(t) & = \mathbf{s}_{r}(t) + \mathbf{\Tilde{x}}_{r}(t) + \mathbf{v}(t) \quad \in \mathbb{R}^{M\times1}
\label{eq:mic}
\end{aligned}
\end{equation}

\noindent where, $\mathbf{s}_{r}(t) = \mathbf{h}_{s}(t)\ast s(t)$ and $\mathbf{\Tilde{x}}_{r}(t) = \mathbf{h}_{x}(t) \ast f_{\text{NL}}(x(t))$ are the received reverberant copies of near-end and audio device emitted nonlinearly distorted far-end speech components, $f_{\text{NL}}$ is a function that mimics nonlinearities introduced by the audio device, $\mathbf{h}_{s}(t)$ and $\mathbf{h}_{x}(t)$ are $M$-channel room impulse responses (RIRs) from near-end speaker and the loudspeaker locations to the microphone array, `$\ast$' denotes the convolution, and $\mathbf{v}(t)$ represents the background noise.

This study presents a deep learning-based joint AEC and beamforming model $(\mathlarger{\boldsymbol \Psi})$ as a supervised speech enhancement system. The proposed model aims to suppress far-end echo with nonlinear distortions and background noise while preserving near-end speech using a mixture and clean far-end signals. As stated in Eq.\eqref{eq:model}, the proposed network is limited to estimating reverberant near-end speech, $\hat{s}_{r}(t)$ but not the anechoic near-end speech, $\hat{s}(t)$. 

\begin{equation} 
\begin{aligned}[b]
\hat{s}_{r}(t) = \boldsymbol \Psi(\mathbf{y}(t)) = \boldsymbol \Psi([\mathbf{d}(t),\; x(t)]^T)
\label{eq:model}
\end{aligned}
\end{equation}


\section{Proposed Method}
\label{sec:proposed_network}
\begin{figure}[ht]
  \centering
  \includegraphics[width=\linewidth]{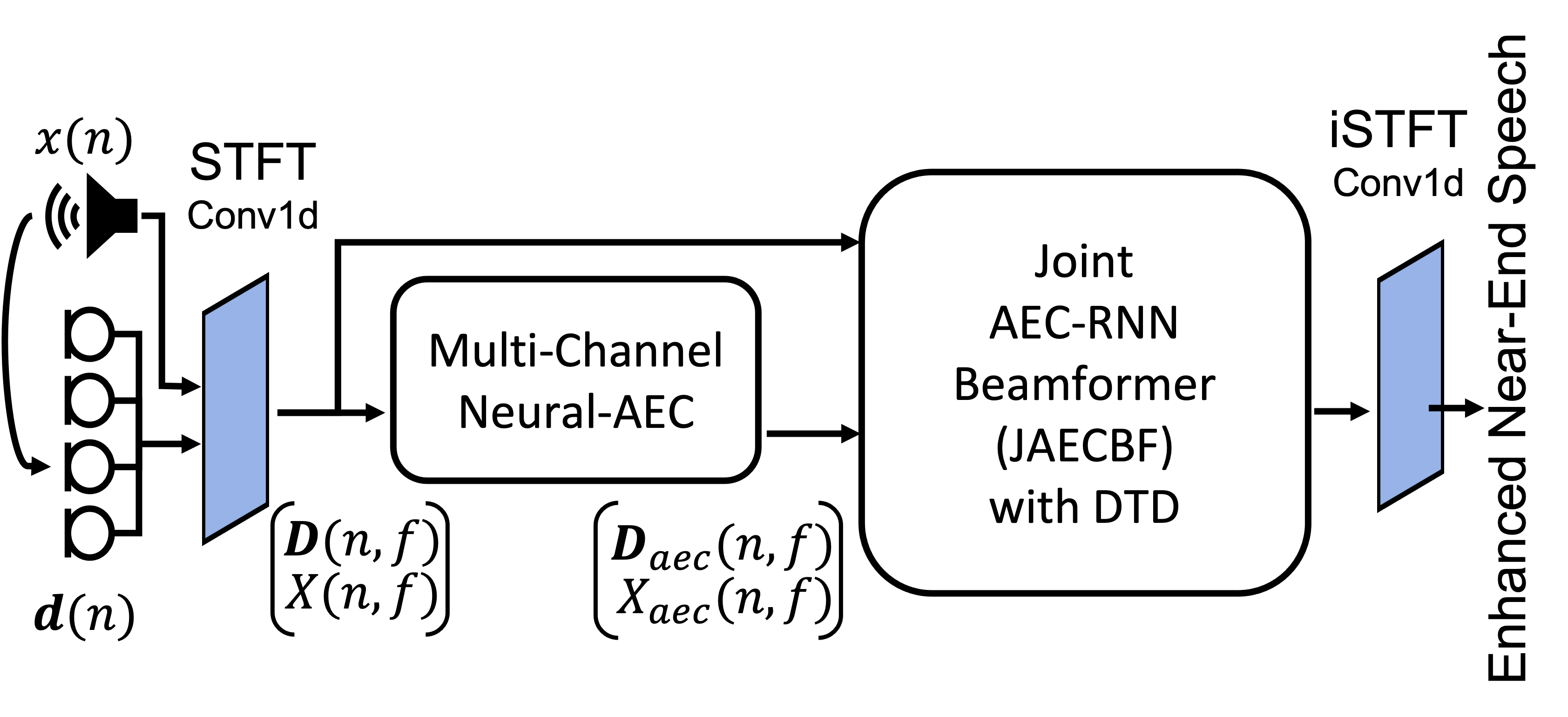}
   \caption{Overview of the proposed joint AEC and beamformer.}
  \label{fig:proposed}
\end{figure}

The overall architecture of the proposed network is depicted in Figure \ref{fig:proposed} which comprises of two modules: (i) a multi-channel neural-AEC (see sec.\ref{sec:aec}), and (ii) a spatio-temporal joint AEC-RNN beamformer with double-talk detection (DTD) module trained using cross-correlations as input features (see sec.\ref{sec:beamformer}). We jointly train these modules in an end-to-end approach. As stated in Eq.\eqref{eq:model}, the proposed model is provided with stacked $M$-channel mixture and single-channel far-end signals $\mathbf{y}(t){\in}\mathbb{R}^{(M{+}1)\times T}$, where `$T$' represents the number of samples. The audio samples are first transformed to frequency domain, $\mathbf{Y}(n,f)$ using one-dimensional convolution layers that employ a short-time fourier transform (STFT) operation. Here, `$n$'${\in}[0,N)$, and `$f$'${\in}[0,F)$ represents frame index and frequency bin. To this end, we use $\mathbf{Y}(n,f)$ to compute TF-bin wise cross-correlations between the mixture and far-end signals `$\mathrm{R}_{\mathbf{Y}\mathbf{Y}}$', to server as input features to our multi-channel neural-AEC. The frame-wise cross-correlations are calculated as:
\begin{equation} 
\begin{aligned}[b]
\mathrm{R}_{\mathbf{Y}\mathbf{Y}}(n,f) =  \big(\mathbf{Y}(n,f)\mathbf{Y}^H(n,f)\big)
\label{eq:spatial_features}
\end{aligned}
\end{equation}
These input features include phase delays between far-end signals and mixture in addition to inter-microphone phase delays that are crucial for engineering a robust multi-channel neural-AEC. To maximize the learning potential, multi-head self-attention (MHSA) \cite{yong1_ATT} over time is employed on input features to dynamically emphasize on relevant correlations.
\vspace{-1em}
\vspace{-1em}
  
\subsection{Multi-channel Neural-AEC}
\label{sec:aec}
\begin{figure}[ht]
  \centering
  \includegraphics[width=\linewidth]{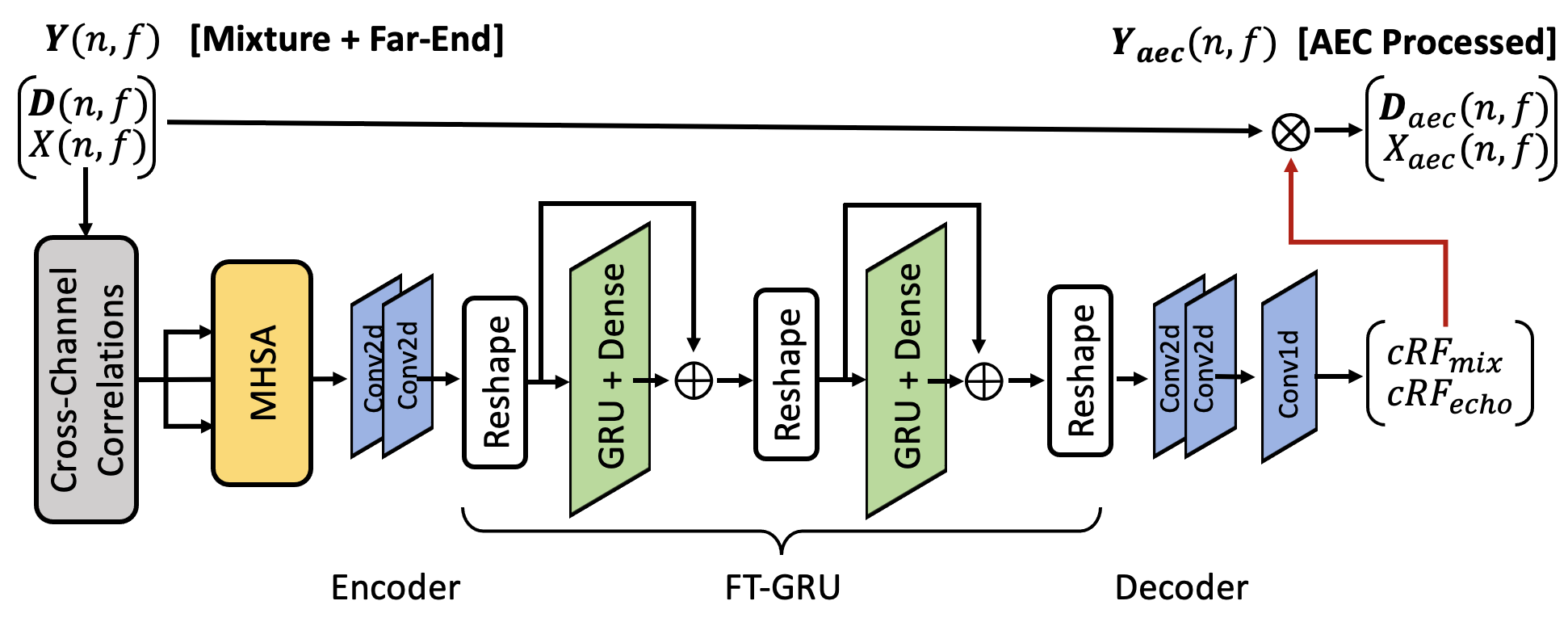}
   \caption{Network architecture of multi-channel neural-AEC.}
  \label{fig:mc_neural_aec}
  \vspace{-1em}
\end{figure}

We design our multi-channel neural-AEC module using a deep convolutional recurrent neural network (DCRNN) \cite{DCRNN} with two encoders, two decoders, and a frequency-time gated recurrent units (FT-GRU) with residual connections that estimates complex-valued ratio filters (cRF) \cite{deepfilters} for mixture and far-end signals, respectively. Figure \ref{fig:mc_neural_aec}, presents the architectural details of the multi-channel neural-AEC module. The encoders within the module extract spatial-temporal information $U_\mathrm{enc}$ from the input features using two-dimensional convolutional layers, which are subsequently fed to FT-GRU.

Similar to \cite{FTLSTM,FTLSTM_original}, we design a frequency-time recurrent network (FT-GRU) by integrating two gated recurrent units (GRU), fully connected (FC) networks, and residual connections. The first GRU network scans all frequency bins to summarize spectral information from encoded features, $U_\mathrm{enc}$. Later, we restructure the output layer's activations and input them into the second GRU network, which analyses correlations over time to produce $U_\mathrm{out}$, see Eq.\eqref{eq:ftgru}. 
\begin{equation}
\begin{split}
    \textrm{FT-GRU} \Bigg\lbrace
\end{split}\:
\begin{split}
    &Z{=}\big(U_{in}{+} \mathrm{FC}\big(\mathrm{GRU}\big(U_\mathrm{enc}[:,f,n]\big)\big)\big)^\mathrm{Tr}\\ \vspace{-2em}
    &U_\mathrm{out}{=}\big(Z{+} \mathrm{FC}\big(\mathrm{GRU}\big(Z[:,n,f]\big)\big)\big)^\mathrm{Tr}
\end{split}
\label{eq:ftgru}
\end{equation}

\noindent Here $(\cdot)^\mathrm{Tr}$ represents transpose operation performed on time and frequency dimensions, and `$Z$' is the intermediate output from the first GRU network. The decoders within the module estimate two $((2K{+}1){\times}(2L{+}1))$-dimensional complex-valued ratio filters ($\mathrm{cRF_{mix}}, \mathrm{cRF_{echo}}$) which are used to process mixture and far-end signals, respectively. As demonstrated in Eq.\eqref{eq:aec_crf}, we employ $\mathrm{cRF_{mix}}$ on multi-channel mixture signals $\mathbf{D}(n,f)$ to produce far-end echo suppressed mixture signals, $\mathbf{D}_\mathrm{aec}(n,f)$. Similar computations are carried on single-channel far-end signal $X(n,f)$ using $\mathrm{cRF_{echo}}$ to produce $X_\mathrm{aec}(n,f)$ that is time-aligned with the enhanced mixture signals. 
\begin{equation}
    \resizebox{0.88\linewidth}{!}{$
    \begin{aligned}
        \mathbf{D}_\mathrm{aec}(n,f)=\smashoperator{\sum_{\tau_1 \in [-K,K],\tau_2 \in [-L,L]}}\mathrm{cRF_{mix}}(n,f,\tau_1,\tau_2)\ast\mathbf{D}(n{+}\tau_1,f{+}\tau_2)
    \end{aligned} 
    $}
    \label{eq:aec_crf}
\end{equation}

\vspace{-1.5em}
\begin{equation} 
\begin{aligned}[b]
\mathbf{\widetilde{Y}}(n,f)= [\mathbf{Y}(n,f), \: \mathbf{D}_\mathrm{aec}(n,f), \: X_\mathrm{aec}(n,f)]^T
\label{eq:aec_output}
\end{aligned}
\end{equation}

Later, the processed mixture and far-end signals are stacked channel-wise with the original signals (termed as ``stacked inputs''), $\mathbf{\widetilde{Y}}(n,f)$, and fed to our beamforming module. 


\subsection{Joint AEC-RNN Beamforer (JAECBF) with DTD}
\label{sec:beamformer}
\begin{figure}[ht]
  \centering
  \includegraphics[width=\linewidth]{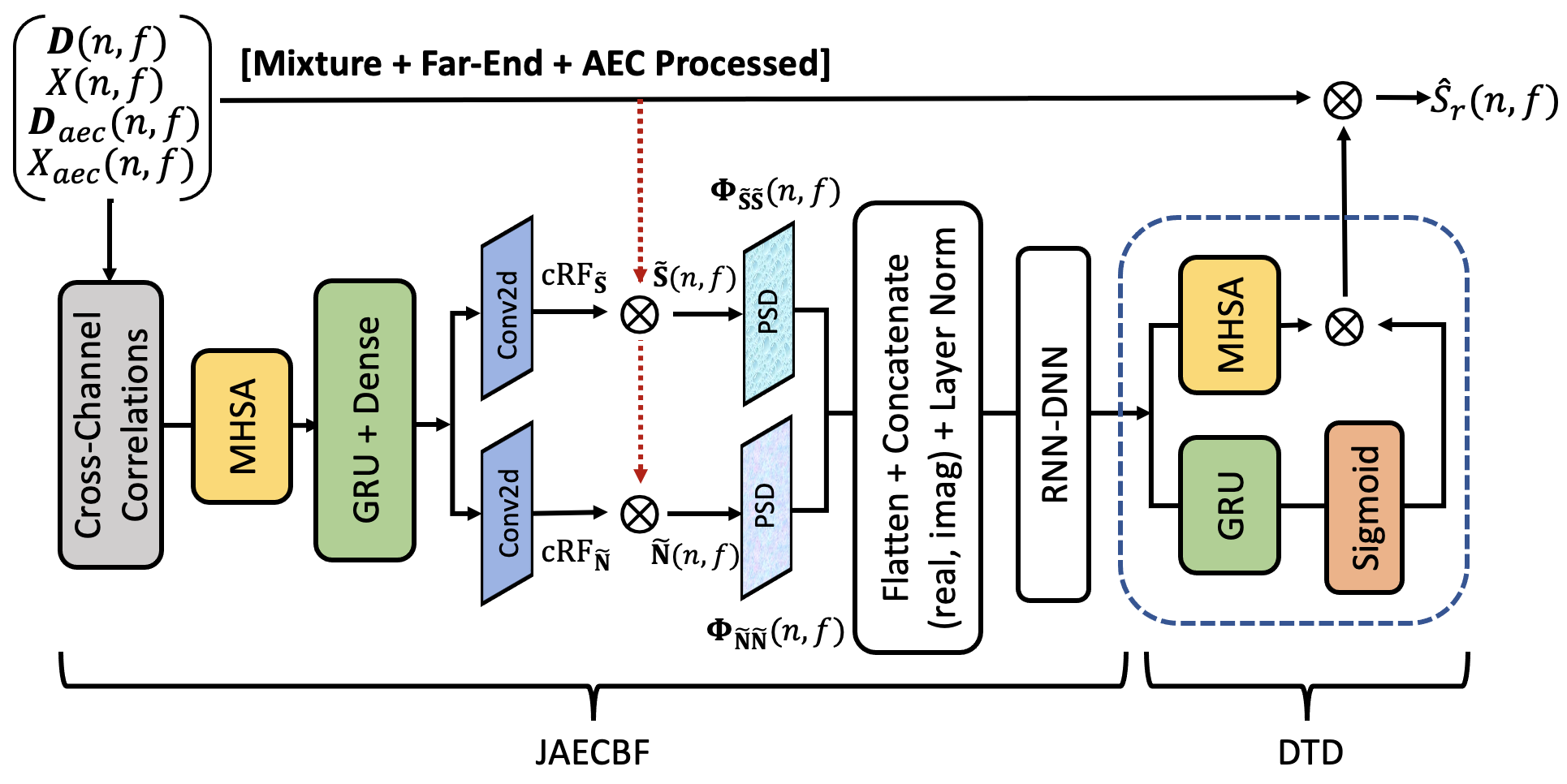}
   \caption{Network architecture of the proposed Joint AEC-RNN Beamformer (JAECBF) with DTD.}
  \label{fig:rnn_beamformer}
  \vspace{-1em}
\end{figure}

The proposed joint AEC-RNN beamformer with double-talk detection (JAECBF) is an extension to our previous work generalized spatio-temporal RNN beamformer (GRNNBF) \cite{rnnbf}. Unlike conventional AEC-beamforming strategies that perform spatial filtering on AEC-processed signals, our JAECBF is designed to learn TF-bin wise beamforming weights to optimally combine the mixture, far-end, and AEC-processed signals to extract near-end speech signal. Figure \ref{fig:rnn_beamformer}, presents the architectural overview of our proposed JAECBF with double-talk-detection. Similar to our multi-channel neural-AEC module, we compute frame-wise cross-channel correlations `$\mathrm{R}_{\mathbf{\widetilde{Y}}\mathbf{\widetilde{Y}}}$' using the stacked inputs $\mathbf{\widetilde{Y}}(n,f)$ to server as input features to the beamforming module. Our beamformer module benefits from the correlations of time-aligned far-end and echo-suppressed mixture signals with the original mixture signals, in addition to the inter-microphone correlations. We use multi-head self-attention (MHSA) over time on cross-channel correlations features to dynamically emphasize on relevant correlations.

These dynamic features from the stacked inputs are then passed through to series of GRU, fully-connected, and one-dimensional convolution layers to estimate two $((2K{+}1){\times}(2L{+}1))$-dimensional complex ratio filters ($\mathrm{cRF_{\mathbf{\widetilde{S}}}}, \mathrm{cRF_{\mathbf{\widetilde{N}}}}$) which are used to compute multi-channel target speech and noise signals. As demonstrated in Eq.\eqref{eq:speech_est}, we employ $\mathrm{cRF_{\mathbf{\widetilde{S}}}}$ on stacked input signals $\mathbf{\widetilde{Y}}(n,f)$ to produce multi-channel target speech signal, $\mathbf{\widetilde{S}}(n,f)$. Similar computations are carried on stacked input signals using $\mathrm{cRF_{\mathbf{\widetilde{N}}}}$ to produce multi-channel noise  signals, $\mathbf{\widetilde{N}}(n,f)$. Next, we compute frame-wise speech covariance matrix $\mathbf{\Phi_{\widetilde{S}\widetilde{S}}}(n,f)$ from multi-channel speech signals. Similar computations are carried out to compute noise covariance matrix $\mathbf{\Phi_{\widetilde{N}\widetilde{N}}}(n,f)$. Similar to our prior work \cite{rnnbf}, we use layer normalization with learnable affine transforms to replace the the conventional mask normalization, see Eq.\eqref{eq:speechPSD}.
\vspace{-0.2em}
\begin{equation} 
\setlength{\jot}{0.1pt}
\begin{aligned}[b]
\mathbf{\widetilde{S}}(n,f)= \smashoperator{\sum_{\tau_1 \in [-K,K],\tau_2 \in [-L,L]}}\mathrm{cRF}_{\mathbf{\widetilde{S}}}(n,f,\tau_1,\tau_2)\: \ast \: \mathbf{\widetilde{Y}}(n{+}\tau_1,f{+}\tau_2)
\label{eq:speech_est}
\end{aligned}
\end{equation}
\vspace{-0.5em}
\begin{equation} 
\setlength{\jot}{0.1pt}
\begin{aligned}[b]
\mathbf{\Phi_{\widetilde{S}\widetilde{S}}}(n,f)= \mathrm{LayerNorm}\big(\mathbf{\widetilde{S}}(n,f)\mathbf{\widetilde{S}}(n,f)^H\big)
\label{eq:speechPSD}
\end{aligned}
\end{equation}
The real and imaginary parts of frame-wise speech and noise covariance matrices. Since $\mathbf{\widetilde{Y}}$ includes the echo reference and the multi-channel signals, $\mathbf{\Phi_{\widetilde{S}\widetilde{S}}}$ and   $\mathbf{\Phi_{\widetilde{N}\widetilde{N}}}$ jointly models the cross-correlation among the echo and the multi-channel signals which are concatenated and fed to a unified RNN-DNN ($\bigF_\textrm{RD}$) to predict beamforming weights for stacked inputs. 

\vspace{-1em}
\begin{equation}
    \resizebox{0.9\linewidth}{!}{$
    \begin{aligned}
        \mathrm{\textbf{W}}_\mathrm{JAECBF}(n,f) = \bigF_\textrm{RD}\big(\big[ \mathbf{\Phi_{\widetilde{S}\widetilde{S}}}(0{:}n,f), \mathbf{\Phi_{\widetilde{N}\widetilde{N}}}(0{:}n,f)\big]\big)
    \end{aligned} 
    $}
    \label{eq:bf_weights}
\end{equation}


\noindent We further adapt JAECBF for AEC applications by adding a double talk detection (DTD) module. The DTD module uses multi-head self attention (MHSA) to compute the global temporal correlations within the TF-bin wise beamforming weights. In addition, it employs a GRU network coupled with sigmoid activation ($\sigma$) which predicts double-talk scenarios. These probabilities are used to scale the JAECBF-estimated beamforming weights accordingly to suppress far-end echoes and coexisting background noise in near-end inactive speech frames.
\vspace{-0.1em}
\begin{equation} 
\setlength{\jot}{0.1pt}
\begin{aligned}[b]
\mathrm{\textbf{W}}_\mathrm{JAECBF{-}DTD} =  \sigma(\textrm{GRU}(w)) \cdot  \textrm{MHSA}(w,w,w)
\label{eq:dtd_weights_1}
\end{aligned}
\end{equation}

\noindent where $w$ represents the complex-valued beamforming weights $\mathrm{\textbf{W}}_\mathrm{JAECBF}(n,f)$. We compute the enhanced near-end speech $\hat{S}_r(n,f)$ using the proposed  beamformer weights, original mixture, far-end, and AEC processed signals. 

\vspace{-1em}
\begin{equation} 
\setlength{\jot}{0.1pt}
\begin{aligned}[b]
\hat{S}_r(n,f) &= (\mathrm{\textbf{W}}_\mathrm{JAECBF-DTD}(n,f)\big)^H\mathbf{\widetilde{Y}}(n,f)
\label{eq:dtd_weights}
\end{aligned}
\vspace{-0.5em}
\end{equation}

Finally, $\hat{S}_r(n,f)$ is transformed to time domain audio samples, $\hat{s}_{r}(t)$ using one-dimensional convolution layers that employ the inverse short-time fourier transform (iSTFT) operation.

\section{Dataset and Experimental Setup}
\label{sec:dataset}
\subsection{Dataset}
\label{ssec:dataset}
We simulate multi-channel reverberant and noisy dataset using AISHELL-2 \cite{AISHELL} and AEC-Challenge \cite{AEC_sim1} corpus. We generate a total of 10k multi-channel RIRs with random room characteristics using image-source method. Each multi-channel RIR is a set consisting of RIRs from near-end speaker, loud-speaker, and background noise locations to 8-channel linear microphone array measuring 26 cm in length. The reverberation time (RT$_{60}$) ranges between [0,0.6s] across room configurations. We randomly select RIRs to simulate multi-channel AEC dataset. We use clean and nonlinear distorted versions of far-end speech from AEC-Challenge \cite{AEC_sim1}. The nonlinear distortions include, but are not limited to: (i) clipping the maximum amplitude, (ii) using a sigmoidal function \cite{AEC_sim2}, and (iii) applying learned distortion functions. In addition, we include diffused noise with SNRs ranging from [0,40] dB and  signal to echo ratio (SER) from [-10,10] dB. A total of 90K, 7.5K, and 2K utterances are generated for the `\textit{Train}', `\textit{Dev}', and `\textit{Test}' datasets.

\vspace{-0.5em}
\subsection{Experimental Setup}
\label{ssec:expsetup}
A 512-point STFT is employed with 32 ms Hann window and 16 ms step size to extract complex spectra for mixture and far-end signals. All systems in the study are trained on 4-second chunks with the Adam optimizer and a batch size of 12 to maximize the time-domain scale-invariant source-to-noise ratio (Si-SNR) \cite{SiSNR} and minimize the frequency-domain mean square error (MSE), both of which are equally weighted. Initial learning rate is set to 1e-4 with a gradient norm clipped with max norm 10. All systems are designed to have $\sim$8.5M parameters and trained over 30 epochs. The estimated cRFs size in the proposed systems is empirically set to (3x1). In this study, we compare our proposed method to four baseline systems which include: (i) SpeexDSP \cite{Speex1}, a purely signal processing-based AEC, (ii)  FT-LSTM \cite{FTLSTM}, an AEC adapted for multi-channel, (iii) GRNNBF \cite{rnnbf}, a robust NN-based beamformer, (iv) hybrid models that combine AEC with various beamformers. The readers can find the script for our proposed method and samples of enhanced audio files at 
\href{https://vkothapally.github.io/JAECBF}{\color{blue}\textup{https://vkothapally.github.io/JAECBF}}.


\begin{table*}[t!]
\centering
\caption{Experimental results for different joint AEC and spatial filtering networks across objective evaluation metrics.}
\label{tab:Results}
\begin{tabular}{l|c|c|c|c|c|c}
    \hline
    \textbf{Systems/Metrics} & \textbf{PESQ ($\uparrow$)} & \textbf{STOI ($\uparrow$)} & \textbf{SISNR ($\uparrow$)} & \textbf{SDR  ($\uparrow$)} & \textbf{ERLE ($\uparrow$)} & \textbf{WER\%  ($\downarrow$)}\\
    \hline
    Reverberant clean reference & 4.500 & 1.000 & $\infty$ & $\infty$ & $\infty$ & 2.190 \\
    Mixture (No Processing) & 1.708 & 0.593 & -4.275 & -3.806 & 0.00 & 77.120 \\ \hline
    SpeexDSP \cite{Speex1} & 1.935 & 0.637 & -1.519 & -0.590 & 3.652 & 44.994 \\ 
    FT-LSTM \cite{FTLSTM} & 2.997 & 0.839 & 10.535 & 11.568 & 33.055 & 15.859 \\ \hline
    GRNNBF \cite{rnnbf} & 2.765 & 0.798 & 9.530 & 10.589 & 34.940 & 23.805 \\
    JRNNBF (GRNNBF w. Far-End) & 2.872 & 0.822 & 10.268 & 11.233 & 34.395 & 17.393 \\ \hline
    SpeexDSP + JAECBF (SP + DL) & 2.811 & 0.810 & 7.212 & 10.849 & 34.100 & 20.465 \\
    FT-LSTM + MVDR (DL + SP)& 2.853 & 0.826 & 7.688 & 9.791 & 34.869 & 13.525 \\
    FT-LSTM + JAECBF (DL + DL) & 3.046 & 0.847 & 11.081 & 11.999 & \textbf{37.420} & 14.028 \\ \hline
    \textbf{Proposed JAECBF w. DTD} & \textbf{3.117} & \textbf{0.858} & \textbf{11.280} & \textbf{12.178} & 36.620 & \textbf{11.392} \\
    \hline
    \multicolumn{7}{c}{$^\ast$SP: Signal Processing algorithm \hspace{1.5cm}$^\ast$DL: Deep Learning model }\\
    \hline
\end{tabular}
\end{table*}

\begin{figure*}[t!]
    \centering
    \includegraphics[width=0.95\textwidth]{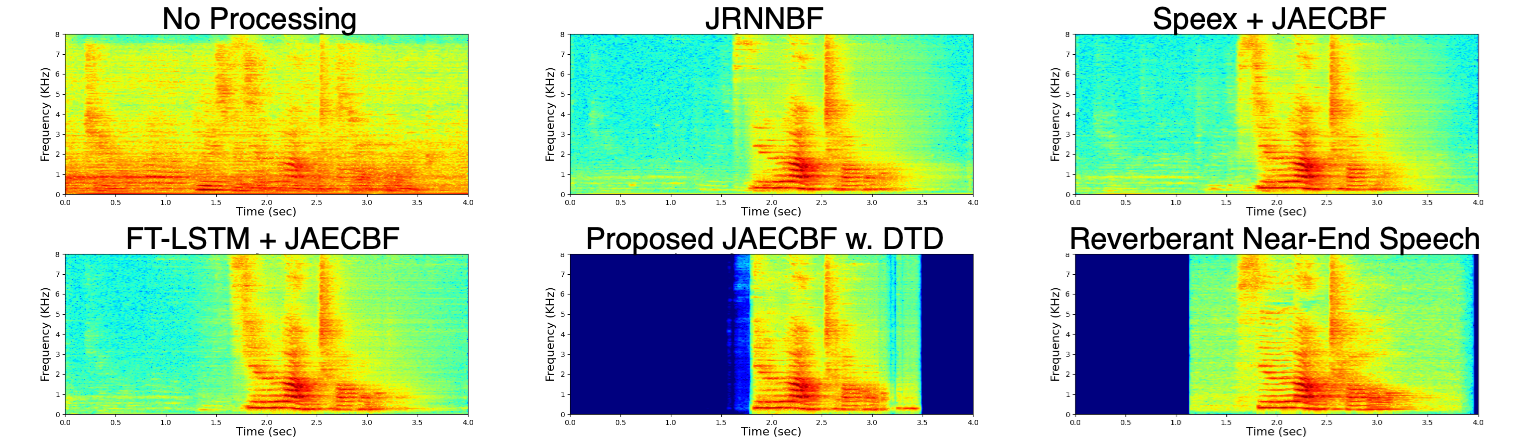}
    \vspace{0.5em}
    \caption{Overview of the proposed joint AEC and beamformer trained using time-domain Si-SNR and frequency domain L2-losses.}
    \label{fig:spectrograms}
    \vspace{-1.3em}
\end{figure*}

\section{Results and Discussions}
\label{sec:results}
We compare the performance of proposed system to other systems using perceptual quality metrics such as PESQ and STOI, as well as objective metrics such as Si-SNR, signal-to-distortion ratio (SDR), and echo return loss enhancement (ERLE) on the `\textit{Test}' set in Table.\ref{tab:Results}. Furthermore, a general-purpose mandarin speech recognition Tencent API \cite{TencentASR} is used to test the ASR performance by computing word error rate (WER). 

\noindent\textbf{[``GRNNBF vs. JRNNBF'']:} GRNNBF learns beamforming weights from mixture and far-end signals to perform spatial filtering using only mixture signals. JRNNBF is an adaptation of GRNNBF, which learns beamforming weights for mixture and far-end signals. As a result, we see that the proposed JRNNBF improves the performance of GRNNBF. For example, an average PESQ of 2.76 vs. 2.87; WER: 23.80 vs. 17.93. Likewise, we also see 3\%, 8\%, and 6\% relative improvements in STOI, Si-SNR, and SDR respectively. These finding suggest that by estimating beamformer weights for far-end signals in addition to mixture, the network learns a better beamforming solution. As a result, we propose JAECBF, which extends JRNNBF to perform spatial filtering using mixture, far-end, and multi-channel AEC processed signals.


\noindent\textbf{[``Hyrbid vs. NN-based'' Joint AEC beamformer]:} To address joint AEC and beamforming, we design different combinations of signal processing and deep learning approaches for AEC and beamforming. First, we build the following two hybrid systems: (i) signal processing (SP)-based SpeexDSP \cite{Speex1} with our proposed JAECBF, and (ii) deep learning (DL)-based FL-LSTM with SP-based minimum variance distortionless response (MVDR) \cite{mvdr}. We compare the performance of these hybrid systems with a NN-based system constructed using FT-LSTM in conjunction with our JAECBF.


From Table-\ref{tab:Results}, we observe that though SpeexDSP has a marginal impact on general speech quality but has a significant effect on the performance of speech recognition systems, i.e., an average PESQ improvement of 1.70 to 1.93 and a corresponding WER improvement of 77.12 to 44.99.
Nonetheless, the hybrid system outperforms SpeexDSP in terms of performance. The performance benefits were not significantly greater than those obtained with our proposed JAECBF beamformer. We hypothesize that the linear adaptive filters in SpeexDSP do not converge because of substantial non-linear distortions in training samples. This increases the range of uncertainty in nonlinear distortions, subsequently lowering the learning ability of JAECBF. The performance degradation can also be observed in Figure \ref{fig:spectrograms}. Likewise, while multi-channel adapted FT-LSTM performs well on its own, it does not perform well when combined with traditional MVDR on speech quality metrics. A probable reason for this is that traditional MVDR prioritizes a distortionless response over suppression to preserve the near-end speech. This can be observed from the improvements achieved in WER: 13.25 vs 20.46 and SiSNR:7.68 vs 10.53 over the hybrid system. However, the adapted FT-LSTM when combined with our proposed JAECBF outperforms both hybrid models. For example, an average PESQ of 3.05 vs. \{2.81,2.85\}; SiSNR: 11.08 vs. \{7.21,7.68\} and ERLE: 37.42 vs \{34.1,34.87\} over hybrid models with the exception on WER which is still in a comparable range. The findings suggest that our proposed JAECBF optimizes well with NN-based AEC systems by including AEC processed signals within alongside mixture and far-end signals.

\noindent\textbf{[``Proposed JAECBF w. DTD vs FT-LSTM + JAECBF'']:} Our proposed joint AEC and beamforming system differs from the ``FT-LSTM + JAECBF'' in the following ways: (i) we replace conventional LPS and IPD input features with cross-channel correlations from Eq.\eqref{eq:spatial_features}, (ii) we use proposed DTD module to scale the estimated beamformer weights to further suppress the far-end echo and background noise in near-end inactive speech regions via double-talk predictions. The proposed system achieves better performance than ``FT-LSTM + JAECBF''  in terms of quality and speech recognition i.e., PESQ: 3.12 vs 3.04, WER: 11.39 vs 14.03. Likewise, we see \{1.7,1.4\}\% relative improvements in SiSNR and SNR respectively. Figure  \ref{fig:spectrograms} also shows that the proposed system can enhance the spectrogram with less residual echo compared to other systems. Although``FT-LSTM + JAECBF" achieves a bit higher ERLE than the proposed, 37.42 vs 36.62, we can conclude that the major contribution to this improvement comes from our proposed JAECBF when compared to "FT-LSTM", 37.42 vs 33.05.

\vspace{-1.2em}
\section{Conclusion}
\label{sec:conclusion}
\vspace{-0.5em}
To conclude, we present an all-deep learning strategy for joint AEC and beamforming with the following major contributions. First, we propose using cross-correlation coupled with multi-head self-attention to learn significant features for AEC. Second, we propose JAECBF, an extension of our previous work GRNNBF, which performs beamforming using mixture, far-end, and AEC processed signals. Finally, we propose DTD module that predicts double-talk using RNN and scales the beamforming weights to compensate for double-talk scenarios. Among systems evaluated, the proposed system achieves the highest objective scores and the lowest WER. Although the proposed system performs well in terms of recognition and echo suppression, we believe addressing dereverberation alongside AEC and beamforming can further improve performance.

\bibliographystyle{IEEEtran}

\bibliography{Main}


\end{document}